\documentclass[11pt, a4paper]{article}
\usepackage[top=1in, bottom=1in, left=1.25in, right=1.25in]{geometry}
\usepackage{booktabs}
\usepackage{multirow}
\usepackage{setspace}
\usepackage{natbib}
\usepackage{authblk}
\usepackage[pdftex]{graphicx}
\usepackage{amsmath}
\usepackage{amsthm}
\usepackage{mathtools}
\usepackage{amsfonts}
\usepackage{ascmac}
\usepackage{amssymb}
\usepackage{bm}
\usepackage{color}
\usepackage{enumitem}
\usepackage{mathrsfs}
\usepackage[ 
	colorlinks  = true, 
	anchorcolor = black,
	linkcolor   = blue,
	citecolor   = blue, 
	filecolor   = black, 
	urlcolor    = black
]{hyperref}
\usepackage{comment}
\usepackage{here}
\usepackage{rotating}
\usepackage{xcolor}
\usepackage{arydshln}
\usepackage{threeparttable}

\newcommand{\biblist}{\begin{list}{}
{\listparindent 0.0cm \leftmargin 0.50cm \itemindent -0.50 cm
\labelwidth 0 cm \labelsep 0.50 cm
\usecounter{list}}\clubpenalty4000\widowpenalty4000}
\newcommand{\ebiblist}{\end{list}}

 \theoremstyle{plain}

\theoremstyle{definition}

\theoremstyle{remark}
\newtheorem*{rem}{Remark}

\title{\bf On window mean survival time with interval-censored data}
\author[1]{Takuto Iijima}
\author[2]{Tomotaka Momozaki}
\author[2]{Shuji Ando}
\affil[1]{Department of  Information Sciences, Graduate School of Science and Technology, Tokyo University of Science}
\affil[2]{Department of  Information Sciences, Faculty of Science and Technology, Tokyo University of Science}

\date{Last update: \today}

\begin{document}
\maketitle
\begin{abstract}
In recent years, cancer clinical trials have increasingly encountered non proportional hazards (NPH) scenarios, particularly with the emergence of immunotherapy. 
In randomized controlled trials comparing immunotherapy with conventional chemotherapy or placebo, late difference and early crossing survivals scenarios are commonly observed. 
In such cases, window mean survival time (WMST), the area under the survival curve within a pre-specified interval $[\tau_0, \tau_1]$, has gained increasing attention due to its superior power compared to restricted mean survival time (RMST), the area under the survival curve up to a pre-specified time point. 
Considering the increasing use of progression-free survival as a co-primary endpoint alongside overall survival, there is a critical need to establish a WMST estimation method for interval-censored data; however, sufficient research has yet to be conducted. 
To bridge this gap, this study proposes a WMST inference method utilizing one-point imputations and Turnbull’s method. 
Extensive numerical simulations demonstrate that the WMST estimation method using mid-point imputation for interval-censored data exhibits comparable performance to that using Turnbull’s method. 
Since the former facilitates standard error calculation, we adopt it as the standard method. 
Numerical simulations on two-sample tests confirm that the proposed WMST testing method have higher power than RMST in late difference and early crossing survival scenarios, while having compatible power to the log-rank test under the PH. 
Furthermore, even when pre-specified $\tau_0$ deviated from the clinically desirable time point, WMST consistently maintains higher power than RMST in late difference and early crossing survivals scenarios. 
\end{abstract}


\medskip

\section{Introduction} \label{sec:intro}
In randomized controlled trials (RCTs) where the outcome is survival time, the log-rank test has been widely used to evaluate differences in survival times between two groups. 
The log-rank test is uniformly most powerful when the proportional hazards (PH) assumption holds. 
However, \cite{uno2014moving} have pointed out that the log-rank test is closely related to the PH assumption, and when the PH assumption is clearly violated, may lack sufficient power to detect differences in survival times between two groups. 
Furthermore, an empirical analysis by \cite{horiguchi2020empirical}, which compare the log-rank test with other testing methods, including restricted mean survival time (RMST) and weighted log-rank tests, using data from recent Phase III cancer RCTs (69 trials for overall survival (OS) and 59 trials for progression-free survival (PFS)), finds that the log-rank test do not always exhibit superior empirical power. 
One potential reason for this result is the inclusion of trials in which the PH assumption does not hold. 
In fact, approximately 7\% of the OS trials and approximately 33\% of the PFS trials violated the PH assumption. 
In subgroup analyses focusing on trials where the PH assumption holds and PFS is the primary endpoint, the log-rank test exhibits the highest empirical power. 
Additionally, a study by \cite{trinquart2016comparison} reports that, as of 2016, 24\% of oncology studies exhibited non-proportional hazards (NPH), indicating an increasing number of cases where routine application of the log-rank test may not be appropriate. 

Based on these facts, alternative methods to the log-rank test that do not rely on the assumption have been developed in recent years. 
Among them, RMST, the area under the survival curve up to a pre-specified time point, is preferred due to its ease of clinical interpretation compared to existing statistical testing methods. 
Discussions on RMST have begun to appear in medical journals such as the \textit{``Journal of Clinical Oncology''}\citep{trinquart2016comparison}, and RMST has already been used as a summary measure for primary endpoints in some studies \citep{song2019matching, guimaraes2020rivaroxaban, qin2023nimotuzumab, birtle2024improved}. 
Although RMST is robust and interpretable even in NPH scenarios, it is not without its drawbacks. 
A reported limitation is its reduced power in cases where survival curves diverge in the late phase of a trial (late difference) or cross in the early phase (early crossing survivals) \citep{zhang2020restricted}. 

A notable example of scenarios leading to late difference or early crossing survivals is cancer immunotherapy, which has attracted significant attention in recent years. 
Due to its therapeutic mechanisms, many trials involving cancer immunotherapy, particularly when comparing with placebo or chemotherapy groups, have observed late differences or early crossing survivals \citep{Kato2023first, wakelee2023perioperative}. 
Furthermore, the immuno-oncology drug development pipeline has expanded rapidly, increasing approximately 2.5 times between 2017 and 2020 \citep{upadhaya2020immuno}, highlighting the growing need for testing methods capable of maintaining power in late difference and early crossing survivals scenarios. 
In this context, window mean survival time \citep[WMST;][]{paukner2021window}, which is equivalent to long-term RMST \citep{zhao2012utilizing, horiguchi2018quantification, vivot2019use}, has been proposed as a summary measure, maintaining the interpretability of RMST, outperforms RMST in cases involving late difference and early crossing survivals. 
As the name suggests, WMST corresponds to the area under the survival curve within a clinically relevant window of time. 
Unlike RMST, WMST allows clinicians to construct a window of time during the later phase of a study when late difference or early crossing survivals are anticipated. 
This enables focused analysis to determine whether the new treatment improves the mean survival time within that specific window of time. 
In fact, numerical simulations by \cite{paukner2021window} demonstrate that WMST achieve higher statistical power than RMST in late difference and early crossing survivals scenarios with right-censored data. 

In Phase III trials targeting advanced cancer, PFS, defined as the time to disease progression or death (whichever occurs first), is increasingly being used not only as a secondary endpoint but also as a primary endpoint \citep{schachter2017pembrolizumab, gadgeel2020updated, sun2021pembrolizumab}. 
For solid tumors, progression is defined by the revised RECIST guidelines \citep{eisenhauer2009new} as tumor growth exceeding a certain threshold or the appearance of new lesions. 
Since progression is identified only by a medical test, such as CT, the exact time point of the progression event is unknown.
Instead, it is known that the progression event occurred within the interval $(l,r]$, where is the time of $l$ is the last examination without progression, and $r$ is the time when progression is confirmed.
Data representing the time to an event in interval form are referred to as interval-censored data. 
Kaplan-Meier (KM) method \citep{kaplan1958nonparametric} cannot directly estimate survival functions using interval-censored data. 
To address this, imputation methods such as right-point imputation or mid-point imputation are used to approximate interval-censored observations as single time points, allowing KM methods to estimate survival functions. 
Additionally, Turnbull's method has been developed as a non-parametric approach to estimate survival functions directly from interval-censored data \citep{turnbull1976empirical}. 
Although right-point imputation is frequently used in oncology and other fields, its statistical justification remains unclear. 
\cite{nishikawa2004behavior} compare the exactness of the survival rate estimation at different times using the three methods introduced above. 
They find that for estimating PFS rates, the method with the smallest mean squared error (MSE) at the majority of time points is KM estimation following mid-point imputation. 
However, no studies have compared these methods for estimating the accuracy of WMST. 

\cite{zhang2020restricted} discuss an estimator for RMST for interval-censored data. 
In their approach, the survival function is estimated using Turnbull's method, and the area under the survival curve is calculated using the linear smoothing method. 
However, this method cannot express the standard error in a closed form, requiring the use of perturbation-resampling methods like the bootstrap method to calculate the standard error, making the computation less straightforward. 
On the other hand, single-point imputation methods, including right-point imputation or mid-point imputation, can express the standard error in a closed form, allowing for easier computation. 
In this study, we evaluate the performance of three methods for WMST estimation designed for interval-censored data: (1) KM estimation following mid-point imputation (mid-point $+$ KM), (2) KM estimation following right-point imputation (right-point $+$ KM), and (3) Turnbull’s estimation. 
Based on the results of the comparison, the most effective method is used to construct a class of WMST estimators and tests. 
Furthermore, this test method is compared with existing testing methods such as RMST test, the log-rank test, and the Fleming-Harrington test (FH test). 

The remainder of this article is organized as follows. 
Section \ref{sec2} describes the definitions of RMST and WMST, estimation methods and the testing procedure for WMST considering interval-censored data. 
Section \ref{sec3} conducts numerical experiments to perform one-sample estimation and assess the exactness of the estimation. 
Subsequently, two-sample tests are conducted to compare WMST with existing testing methods such as RMST, weighted log-rank tests, and others. 
Additionally, under survival scenarios considering the value of $\tau_0$ (set prior to data collection) and the difference in the areas under two survival curves, the detection power of WMST is compared and evaluated against existing testing methods, including RMST, the log-rank and FH tests. 
Section \ref{sec4} applies WMST test to real-world data and compares it with existing testing methods, as in Section \ref{sec3}. 
Finally, Section \ref{sec5} provides the conclusions.

\section{WMST for interval-censored data} \label{sec2}
In this section, RMST and WMST, which are interpretable even under NPH, are defined along with their estimation methods. 
Hypothesis testing using WMST is also explained. 
Furthermore, the estimation methods designed for interval-censored data in this study are described. 

\subsection{Estimation of RMST for right-censored data}
Let $S(t)$ denote the survival probability at time $t \geq 0$. 
RMST of survival time $T$ is defined as the area under the survival curve up to a specific time point $\tau_1$ \citep{royston2013restricted}, that is, 
\begin{equation*}
    \mathrm{RMST}(\tau_1) = \int^{\tau_1}_0S(t) dt.
\end{equation*}
Let $S_0(t)$ and $S_1(t)$ denote the survival probabilities for the control group and the treatment group, respectively. 
The difference in RMST between the two groups is expressed as 
\begin{align*}
    \delta(\tau_1)
    &=\int^{\tau_1}_0S_1(t)dt - \int^{\tau_1}_0S_0(t)dt\\
    &= \int^{\tau_1}_0\left(S_1(t)-S_0(t)\right)dt.
\end{align*}
With observed $k$ ordered event times denoted as $t_1 < \cdots < t_k$, the estimators of $\mathrm{RMST}(\tau_1)$ and $\delta(\tau_1)$ can be obtained as
\begin{align*}
    \widehat{\mathrm{RMST}}(\tau_1) &= \sum_{t_i\leq\tau_1}(t_{i+1}^*-t_i)\hat{S}(t_i), \\
    \widehat{\delta}(\tau_1) &= \sum_{t_{1,i}\leq\tau_1}(t_{1,i+1}^*-t_{1,i})\hat{S}_1(t_{1,i}) - \sum_{t_{0,i}\leq\tau_1}(t_{0,i+1}^*-t_{0,i})\hat{S}_0(t_{0,i}),
\end{align*}
respectively, where $t_{0,i}$ and $t_{1,i}$ are the ordered $i$-th event times in each group, $t^*_{i+1} = \min(t_{i+1}, \tau_1)$, $t^*_{j,i+1} = \min(t_{j,i+1}, \tau_1)$ ($j \in \{0,1\}$).
$\hat{S}_0(\cdot)$ and $\hat{S}_1(\cdot)$ denote the survival functions estimated by KM method for each group. 

Survival time is often collected as right-censored data, where the event time is known to exceed a certain time because continuous observation until the event occurs is not always feasible. 
For right-censored data, KM method is widely used as a nonparametric approach to estimate the survival function. 

\subsection{Estimation of WMST for right-censored data}
WMST of survival time $T$ is defined as the area under the survival function $S(t)$ from $\tau_0$ to $\tau_1$ \citep{paukner2021window}, that is, 
\begin{equation*}
    \mathrm{WMST}(\tau_0, \tau_1) = \int^{\tau_1}_{\tau_0}S(t) dt.
\end{equation*}
When $\tau_0=0$, WMST is equivalent to RMST. The difference in WMST between the two groups is expressed as 
\begin{align*}
    \Delta(\tau_0, \tau_1)
    &=\int^{\tau_1}_{\tau_0}S_1(t)dt - \int^{\tau_1}_{\tau_0}S_0(t)dt\\
    &= \int^{\tau_1}_{\tau_0}S_1(t)-S_0(t)dt.
\end{align*}
$\mathrm{WMST}(\tau_0,\tau_1)$ and $\Delta(\tau_0, \tau_1)$ can be estimated as 
\begin{align*}
    \widehat{\mathrm{WMST}}(\tau_0, \tau_1) 
    &= \sum_{t_i\in\{\tau_0, \tau_1\}}(t_{i+1}^*-t_i^*)\hat{S}(t_i^*), \\
    \widehat{\Delta}(\tau_0, \tau_1) 
    &= \sum_{t_{1,i}\in\{\tau_0, \tau_1\}}(t_{1,i+1}^*-t_{1,i}^*)\hat{S}(t_{1,i}^*) - \sum_{t_{0,i}\in\{\tau_0, \tau_1\}}(t_{0,i+1}^*-t_{0,i}^*)\hat{S}(t_{0,i}^*), 
\end{align*}
respectively, where $t^*_i = \max(t_i,\tau_0)$ and $t^*_{j,i} = \max(t_{j,i},\tau_0)$ ($j \in \{0,1\}$). 

\begin{rem}
For selecting $\tau_1$, several discussions have been presented in the context of RMST and long-term RMST \citep{horiguchi2018flexible, eaton2020designing, hasegawa2020restricted, tian2020empirical}. 
\cite{tian2020empirical} propose a condition for empirically selecting $\tau_1$ as the smaller of the maximum observed times between the two groups. 
For selecting $\tau_0$, methods have been proposed to select the time point at which the two survival curves begin to diverge or the time at which WMST difference becomes significantly different from zero \citep{horiguchi2023assessing}. 
In this study, $\tau_0$ is set to a predefined value based on the anticipated survival scenario prior to the start of the trial, and $\tau_1$ is selected as the smaller of the maximum observed times between the two groups for hypothesis testing. 
\end{rem}

\subsection{Hypothesis testing with WMST}
The estimator for the variance of WMST can be obtained as 
\begin{align*}
    \hat{\sigma}^2 
    =& \displaystyle\bm\sum_{t_i\in\{\tau_0,\tau_1\}}(t^*_{i+1}-t^*_i)^2\widehat{\mathrm{var}}\left[\hat{S}(t^*_i)\right] \\ 
    &+ \displaystyle\bm\sum_{t_i\in\{\tau_0,\tau_1\}}\displaystyle\bm\sum_{t_j<t_i}(t^*_{i+1}-t^*_i)(t^*_{j+1}-t^*_j)\widehat{\mathrm{cov}}\left[\hat{S}(t^*_i),\hat{S}(t^*_j)\right],
\end{align*}
where 
\begin{align*}
    \widehat{\mathrm{var}}[\hat{S}(t)] &\approx \hat{S}^2(t)\displaystyle\bm\sum_{k|t_k<t}\frac{d_k}{r_k(r_k-d_k)}, \\
    \widehat{\mathrm{cov}}[\hat{S}(t_i),\hat{S}(t_j)] &\approx \hat{S}(t_i)\hat{S}(t_j)\displaystyle\bm\sum_{k|t_k<\min(t_i,t_j)}\frac{d_k}{r_k(r_k-d_k)}, 
\end{align*}
derived using Greenwood's formula, with the number of events $d_k$ and the number of individuals at risk $r_k$ (those in the risk set) at time $t = t_k$. 
Therefore, under the assumption that $S_0(t)$ and $S_1(t)$ are independent, the estimator for the variance of the difference in WMST, $\hat{\Delta}(\tau_0, \tau_1)$, can be obtained as
\begin{equation*}
    \hat{\sigma}^2_{\Delta}=\widehat{\mathrm{var}}[\hat{\Delta}(\tau_0,\tau_1)]=\hat{\sigma}_0^2+\hat{\sigma}_1^2,
\end{equation*}
where $\hat{\sigma}_0^2$ and $\hat{\sigma}_1^2$ are the estimators for the variance of WMST for each group.

When comparing WMST for the control and treatment groups, the following null hypothesis and alternative hypothesis are used:
\begin{equation*}
    H_0:\Delta(\tau_0, \tau_1) = 0, \quad H_1:\Delta(\tau_0, \tau_1) \neq 0.
\end{equation*}
Under the null hypothesis $H_0$, the following test statistic is given as
\begin{equation*}
    Z = \frac{\hat{\Delta}(\tau_0,\tau_1)}{\hat{\sigma}_\Delta}.
\end{equation*}
This test statistic, due to the asymptotic normality of $Z$, can be evaluated against the standard normal distribution under large-sample conditions.

\subsection{WMST estimate designed for interval censoring}
Since this study involves survival time data with interval censoring, KM method cannot be directly applied. 
To this end, we use right-point or mid-point imputation to handle the interval-censored observations, estimate the survival function using KM method, and then calculate WMST. 
These methods are straightforward to implement using software such as \texttt{SAS} or \texttt{R} and allow the explicit calculation of standard errors for the survival estimates. 
Additionally, we compare these approaches with Turnbull's method \citep{turnbull1976empirical}, which directly estimates the survival function from interval-censored data using non-parametric maximum likelihood estimation (NPMLE) without requiring imputation, and then uses the estimated survival function to calculate WMST. 
Our approach to handling interval censoring is summarized as follows. 
\begin{enumerate}
    \item mid-point $+$ KM: The survival function is estimated with KM method after imputing the interval-censored observations $(l_i, r_i]$ by 
        \begin{equation*}
            t_i=\frac{l_i+r_i}{2}.
        \end{equation*}
    \item right-point $+$ KM: The survival function is estimated with KM method after imputing the interval-censored observations $(l_i, r_i]$ by 
        \begin{equation*}
            t_i=r_i
        \end{equation*}
    \item Turnbull's algorithm: Let $t_i$ $(i = 1, \ldots, n)$ denote survival times that follow the survival function $S(\cdot)$, where the interval-censored observation of $t_i$ is given as the interval $(l_i, r_i]$. 
    Then the likelihood function can be expressed as 
        \begin{equation*}
            L(S)\propto\prod_{i=1}^nPr(t_i\in(l_i,r_i])=\prod_{i=1}^n\left(S(l_i)-S(r_i)\right).
        \end{equation*}
    and the survival function is estimated by maximizing this likelihood. 
\end{enumerate}
Compared to the Turnbull method, the methods combining mid-point and right-point imputations with KM estimation are easier to implement, requiring only simple data preprocessing.

\section{Simulation studies} \label{sec3}
This section begins with a description of the settings common to all numerical simulations.
It then outlines the settings and results used to evaluate estimation accuracy, as well as those for assessing test size and power. 
Finally, to examine the robustness of WMST with interval-censored data, the section investigates the impact of RMST difference $\delta$, the starting point of WMST, $\tau_0$, and the divergence or crossing point of the survival curves $x$ on the tests, focusing on survival scenarios frequently observed in cancer immunotherapy trials, such as early crossing survivals and late difference.

\subsection{Simulation settings}
The interval-censored data used in the simulations are generated following \cite{zhang2020restricted}. 
The procedure for simulating a scenario where interval censoring arises in PFS is described below. 
For each subject, a baseline examination is conducted at time $g_0$, followed by $K$ periodic follow-up examinations $\{g_k = g_0 + k \times 1/(K+1)\}$ at equally spaced intervals. 
Additionally, to account for the practical situation in clinical trials where subjects may miss scheduled visits, a dropout probability vector $\bm{P}_{dropout}$ of length $K$ is introduced. 
It is assumed that all subjects complete the baseline examination. 
Furthermore, to investigate how the proportion of interval censoring in observed events impacts the performance of WMST, a parameter $p_{exact}$ is introduced. 
The following steps are repeated for $i = 1, \dots, n$. 
\begin{enumerate}[label=Step \arabic*., leftmargin=*]
    \item Generate the baseline examination time $g_{0(i)}$ from ${\rm Unif} (0, 1/(K+1))$ and construct the examination schedule $\{ g_{0(i)}, g_{1(i)}, \dots, g_{K(i)} \}$. 
    \item Generate the event time $t_i$ from a specific distribution described later. 
    Additionally, based on the dropout probability vector $\bm{P}_{dropout}$, determine whether the examination is missed at time $g_{j(i)}$ $(j=1, \dots, K)$. 
    Generate a random flag $\xi_i$ from ${\rm Bernoulli}(p_{exact})$ to specify whether the event is death or progression. 
    \item If $\xi_i = 1$, set the observed death event as $l_i = t_i - \epsilon, r_i = t_i$ (where $\epsilon$ is a small positive number close to 0). 
    If $\xi_i = 0$, assign the shortest interval $(g_{k(i)}, g_{l(i)}]$ (where $k < l$) that covers $r_i$ as the observed progression event $(l_i, r_i]$. 
    In this case, both $g_{k(i)}$ and $g_{l(i)}$ must correspond to time points where the examination is not missed. 
    \item If the event occurs after the final examination time $g_{K(i)}$ or if the progression event cannot be confirmed due to missed examinations, set $r_i = \infty$ and treat it as right-censored.
\end{enumerate}

\subsection{Evaluation of estimation}
\begin{figure}[H]
\centering
\includegraphics[width=\columnwidth]{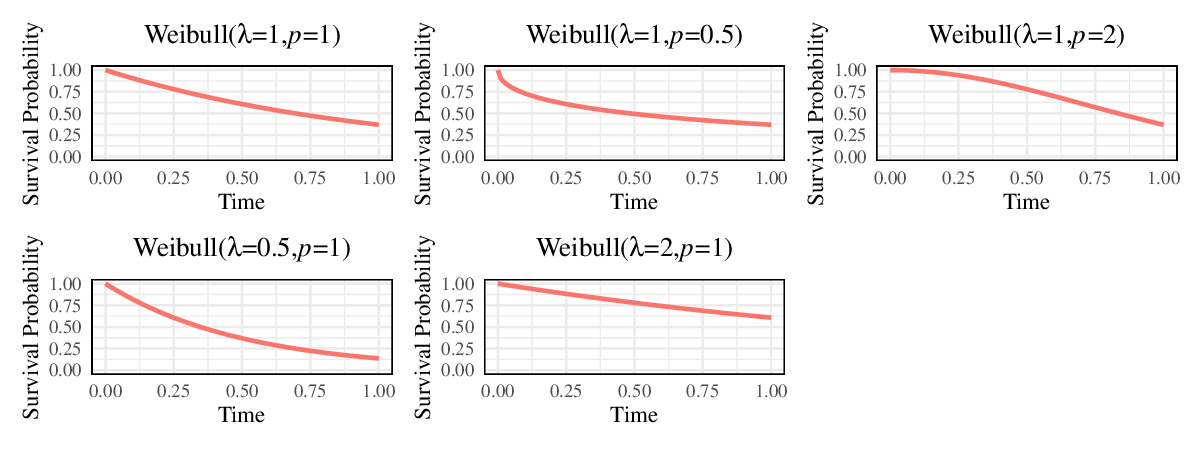}
\caption{Survival curves of five Weibull distributions with different scale parameters ($\lambda$) and shape parameters ($p$) for event-free times.}
\label{fig2}
\end{figure}

In the numerical simulations for evaluating estimation methods, various parameter settings are examined to assess the performance of WMST estimator, taking interval censoring into account. 
First, for the dropout probability vector $\bm{P}_{dropout}$, the following four dropout scenarios are considered. 
In this case, the probability of missing any given examination, except for the final examination, is assumed to be equal across all follow-up examinations, while the dropout probability for the final examination is set to be twice as high. 
Namely, 
\begin{enumerate}[label=(\roman*)]
    \item \textit{None} $:\bm{P}_{dropout,k}=0$ for $k=1,\dots,K$;
    \item \textit{Low} $:\bm{P}_{dropout,k}=0.1$ for $k=1,\dots,K-1$ and $\bm{P}_{dropout,K}=0.2$;
    \item \textit{Medium} $:\bm{P}_{dropout,k}=0.2$ for $k=1,\dots,K-1$ and $\bm{P}_{dropout,K}=0.4$;
    \item \textit{High} $:\bm{P}_{dropout,k}=0.3$ for $k=1,\dots,K-1$ and $\bm{P}_{dropout,K}=0.6$.
\end{enumerate}
Event times have the Weibull distribution with the probability density function
\begin{equation*}
    f_{\rm Weibull}(t) = \frac{pt^{p-1}}{\lambda^p}\exp\left(-\frac{t}{\lambda}^p\right) \quad (\lambda > 0).
\end{equation*}
We considered the five combinations of scale parameter $\lambda$ and shape parameter $p$
\begin{equation*}
    (\lambda,p) \in \{(1, 1), (1, 0.5), (1, 2), (0.5, 1), (2, 1)\}  
\end{equation*}
taking into account survival scenarios where the risk of death is constant, decreasing, or increasing over time, as well as scenarios with high or low mortality risk. 
Figure \ref{fig2} shows the survival curves of the Weibull distributions examined.

The simulation parameters also include sample sizes $n \in \{100, 200, 400\}$, the number of follow-up examinations $K \in \{3, 5, 10, 20\}$, and the proportion of death events among all observed events $p_{exact} \in \{0, 0.2, 0.5, 1.0\}$. 
The numerical simulations are conducted with the event time distribution set to ${\rm Weibull}(\lambda=1, p=1)$, the dropout probability vector $\bm{P}_{dropout,k}$ set to \textit{Medium}, $n=100$, $K=5$, and $p_{exact}=0$ as a default setting. 
The influence of each parameter is evaluated by varying one parameter at a time while keeping the others fixed. 
For each parameter configuration, 10,000 numerical simulations are performed, and the estimation accuracy is examined in terms of relative bias (rBias) and mean squared error (MSE)
\begin{equation*}
    \mathrm{rBias}=\frac{\sum_{s=1}^{10000}\hat{\mu}_s(\tau_0,\tau_1)-\mu(\tau_0,\tau_1)}{\mu(\tau_0,\tau_1)}, \quad 
    \mathrm{MSE}=\frac{1}{10000}\sum_{s=1}^{10000}\left(\hat{\mu}_s(\tau_0,\tau_1)-\mu(\tau_0,\tau_1)\right)^2,
\end{equation*}
where $\hat{\mu}_s(\tau_0,\tau_1)$ is estimated WMST during the $s$-th numerical simulation.

\begin{sidewaystable*}[!htbp]%
\centering
\caption{Simulation results for the estimation of interval-censored WMST with $\tau_0 \in \{0.25,0.50\}$ using three methods:~mid-point imputation$+$KM,~right-point imputation$+$KM and Turnbull method.}
\label{tab1}
\begin{tabular*}{\textwidth}{@{\extracolsep\fill}llrrrrrrrrrrrrrr@{}}
\toprule
&&\multicolumn{6}{c}{$\bm{\tau_0=0.25}$}&&\multicolumn{6}{c}{$\bm{\tau_0=0.50}$}\\ \cmidrule{3-8}\cmidrule{10-15}
&&\multicolumn{2}{c}{\raisebox{-0.1em}{\textbf{mid-point $+$ KM}}}&\multicolumn{2}{c}{\raisebox{-0.1em}{\textbf{right-point $+$ KM}}}&\multicolumn{2}{c}{\raisebox{-0.1em}{\textbf{Turnbull}}}&&\multicolumn{2}{c}{\raisebox{-0.1em}{\textbf{mid-point $+$ KM}}}&\multicolumn{2}{c}{\raisebox{-0.1em}{\textbf{right-point $+$ KM}}}&\multicolumn{2}{c}{\raisebox{-0.1em}{\textbf{Turnbull}}}\\ \cdashline{3-4}[3pt/2pt]\cdashline{5-6}[3pt/2pt]\cdashline{7-8}[3pt/2pt]\cdashline{10-11}[3pt/2pt]\cdashline{12-13}[3pt/2pt]\cdashline{14-15}[3pt/2pt]
\raisebox{-0.3em}{\textbf{Parameter}}&&\raisebox{-0.3em}{\textbf{rBias}}&\raisebox{-0.3em}{\textbf{MSE}}&\raisebox{-0.3em}{\textbf{rBias}}&\raisebox{-0.3em}{\textbf{MSE}}&\raisebox{-0.3em}{\textbf{rBias}}&\raisebox{-0.3em}{\textbf{MSE}}&&\raisebox{-0.3em}{\textbf{rBias}}&\raisebox{-0.3em}{\textbf{MSE}}&\raisebox{-0.3em}{\textbf{rBias}}&\raisebox{-0.3em}{\textbf{MSE}}&\raisebox{-0.3em}{\textbf{rBias}}&\raisebox{-0.3em}{\textbf{MSE}}\\
\midrule
\multicolumn{2}{l}{$T\sim$~Weibull($\lambda$, $p$)}&&&&&&&&&\\ \cmidrule(r){1-2}
Weibull(1,1)$^{\rm a}$&&$-0.0068$&\textbf{0.0011}&0.0949&0.0024&$-0.0065$&\textbf{0.0011}&&$-0.0147$&\textbf{0.0006}&0.0727&0.0008&$-0.0104$&\textbf{0.0006}\\ \hdashline
Weibull(1,0.5)&&$-0.0061$&\textbf{0.0012}&0.0727&0.0018&$-0.0084$&0.0013&&$-0.0149$&\textbf{0.0006}&0.0382&\textbf{0.0006}&$-0.0122$&\textbf{0.0006}\\ \hdashline
Weibull(1,2)&&$-0.0111$&0.0008&0.0848&0.0024&$-0.0038$&\textbf{0.0007}&&$-0.0168$&\textbf{0.0005}&0.0965&0.0012&$-0.0079$&\textbf{0.0005}\\ \hdashline
Weibull(0.5,1)&&$-0.0047$&\textbf{0.0008}&0.2397&0.0039&$-0.0116$&\textbf{0.0008}&&$-0.0243$&\textbf{0.0004}&0.2032&0.0009&$-0.0190$&\textbf{0.0004}\\ \hdashline
Weibull(2,1)&&$-0.0057$&\textbf{0.0008}&0.0393&0.0040&$-0.0046$&0.0009&&$-0.0100$&\textbf{0.0005}&0.0266&\textbf{0.0005}&$-0.0075$&\textbf{0.0005}\\
\midrule
$\bm{P}_{dropout}$&&&&&&&&&&\\ \cmidrule(r{5mm}){1-1}
None&&$-0.0041$&\textbf{0.0010}&0.0739&0.0019&$-0.0053$&0.0011&&$-0.0074$&\textbf{0.0005}&0.0642&0.0008&$-0.0079$&\textbf{0.0005}\\ \hdashline
Low&&$-0.0052$&\textbf{0.0010}&0.0862&0.0022&$-0.0055$&0.0011&&$-0.0102$&\textbf{0.0006}&0.0717&0.0008&$-0.0084$&\textbf{0.0006}\\ \hdashline
High&&$-0.0095$&\textbf{0.0011}&0.0957&0.0024&$-0.0093$&0.0012&&$-0.0211$&\textbf{0.0006}&0.0605&0.0007&$-0.0151$&\textbf{0.0006}\\
\midrule
$n$&&&&&&&&&&\\ \cmidrule(r{5mm}){1-1}
200&&$-0.0026$&\textbf{0.0005}&0.0969&0.0020&$-0.0032$&\textbf{0.0005}&&$-0.0078$&\textbf{0.0003}&0.0759&0.0006&$-0.0051$&\textbf{0.0003}\\ \hdashline
400&&$-0.0006$&\textbf{0.0003}&0.0977&0.0018&$-0.0017$&\textbf{0.0003}&&$-0.0046$&\textbf{0.0001}&0.0771&0.0005&$-0.0027$&\textbf{0.0001}\\
\midrule
$K$&&&&&&&&&&\\ \cmidrule(r{5mm}){1-1}
3&&$-0.0092$&\textbf{0.0011}&0.1111&0.0029&$-0.0097$&0.0012&&$-0.0239$&0.0007&0.0762&0.0009&$-0.0150$&\textbf{0.0006}\\ \hdashline
10&&$-0.0028$&\textbf{0.0010}&0.0602&0.0015&$-0.0026$&\textbf{0.0010}&&$-0.0057$&\textbf{0.0005}&0.0516&0.0006&$-0.0044$&\textbf{0.0005}\\ \hdashline
20&&$-0.0015$&\textbf{0.0010}&0.0329&0.0012&$-0.0014$&\textbf{0.0010}&&$-0.0032$&\textbf{0.0005}&0.0297&0.0006&$-0.0027$&\textbf{0.0005}\\
\midrule
$P_{exact}$&&&&&&&&&&\\ \cmidrule(r{5mm}){1-1}
0.2&&$-0.0132$&\textbf{0.0011}&0.0691&0.0017&$-0.0118$&\textbf{0.0011}&&$-0.0249$&\textbf{0.0006}&0.0466&\textbf{0.0006}&$-0.0198$&\textbf{0.0006}\\ \hdashline
0.5&&$-0.0219$&\textbf{0.0011}&0.0303&\textbf{0.0011}&$-0.0205$&\textbf{0.0011}&&$-0.0386$&0.0006&0.0074&\textbf{0.0005}&$-0.0346$&0.0006\\ \hdashline
1.0&&$-0.0339$&\textbf{0.0012}&$-0.0339$&\textbf{0.0012}&$-0.0339$&\textbf{0.0012}&&$-0.0576$&\textbf{0.0007}&$-0.0576$&\textbf{0.0007}&$-0.0576$&\textbf{0.0007}\\
\bottomrule
\end{tabular*}
\begin{tablenotes}
\item Abbreviation: rBias: relative bias. MSE: mean squared error.
\item $^{\rm a}$ The default simulation setting is the $n=100$, $K=5$, $p_{exact}=0$, $T\sim{\rm Weibull}(1,1)$ and \textit{Medium} dropout rate.
\end{tablenotes}
\end{sidewaystable*}

Table \ref{tab1} shows that the mid-point $+$ KM and Turnbull had a smaller MSE compared to right-point $+$ KM overall. 
However, the difference in MSE between mid-point $+$ KM and Turnbull is negligible. 
When $\tau_0 = 0.50$, the difference in MSE between the mid-point $+$ KM and Turnbull is also negligible, but in terms of the absolute value of rBias, Turnbull outperforms the mid-point $+$ KM. 
Regarding the distribution of survival time $T$, for ${\rm Weibull}(\lambda=1, p=2)$, MSE is nearly identical between the mid-point $+$ KM and Turnbull, but Turnbull shows higher accuracy than the mid-point $+$ KM in terms of the absolute value of rBias. 
Conversely, for ${\rm Weibull}(\lambda=0.5, p=1)$, mid-point $+$ KM outperforms Turnbull in the absolute value of rBias. 

This pattern can be explained by differences in the survival time distributions. 
Among the five distributions examined, ${\rm Weibull}(\lambda=1, p=2)$ is characterized by increasing hazards toward the later stages of the trial. 
Under such conditions, Turnbull demonstrates higher accuracy in terms of the absolute value of rBias. 
To further compare the estimation accuracy of the mid-point $+$ KM and Turnbull under a distribution with even higher hazards in the later stages, ${\rm Weibull}(\lambda=1, p=3)$ is also considered. 
The results show that while the MSE values are almost identical, Turnbull demonstrates higher accuracy in terms of the absolute value of rBias compared to the mid-point $+$ KM. 
In the most notable case among the 10,000 numerical simulations, Turnbull accurately estimates the true survival curve, whereas the mid-point $+$ KM underestimates the survival curve for most of the trial duration, particularly near the end of the trial period, leading to a significant divergence. 
This discrepancy likely contributes to the increase in the absolute value of rBias for the mid-point $+$ KM. 
However, in the field of cancer immunotherapy, where WMST is particularly relevant, survival scenarios such as ${\rm Weibull}(\lambda=1, p=2)$, with increasing hazards in the last part of the trial period, are considered uncommon. 

As $n$ increases, the absolute value of rBias decreases for both the mid-point $+$ KM and Turnbull, while it shows little change for the right-point $+$ KM. 
When the parameter $K$ decreases, the interval width of the interval-censored data increases. 
However, even with small values of $K$, the MSE values for the mid-point $+$ KM and Turnbull remains nearly unchanged. 
Similarly, the parameter $\bm{P}_{dropout}$, which also affects the width of the interval, shows little impact on the MSE values for the mid-point $+$ KM and Turnbull. 
In contrast, for the right-point $+$ KM, the MSE values varies depending on changes in $K$ and $\bm{P}_{dropout}$. 
Furthermore, even with an increased proportion of interval censoring, the MSE values for the mid-point $+$ KM and Turnbull remains nearly constant.

Overall, MSE is nearly equal between the mid-point $+$ KM and Turnbull in most survival scenarios, with the mid-point $+$ KM slightly outperforming Turnbull in many cases. 
Additionally, when estimating the survival function using Turnbull, computational techniques such as bootstrap methods are required to estimate the variance, whereas the mid-point $+$ KM method allows for the explicit expression of the standard error of WMST. 
Therefore, the mid-point $+$ KM is considered to be a more practical and useful method.

\subsection{Evaluation of hypothesis test}
Based on the comparison of the three estimation methods discussed in the previous subsection, the mid-point imputation method, which demonstrates high performance and the ability to explicitly express the standard error, is used to impute interval-censored data. 
The survival function estimated using KM method is then used to calculate WMST, and hypothesis testing based on WMST is evaluated. 
 
\cite{harrington1982class} introduce FH test as a type of weighted log-rank test (WRT). 
The test statistic for WRT is given as
\begin{equation*}
Z_{\rm WRT}=\frac{\sum_{t=1}^D w_t(o_t-e_t)}{\sqrt{\sum_{t=1}^D w_t^2v_t}},
\end{equation*}
where $o_t$ is the observed number of events in the treatment group at time $t$, $e_t$ is the expected number of events under the null hypothesis $H_0$ at time $t$, and $v_t$ is the variance of the number of events in the treatment group under $H_0$. 
The $\hat{S}(t-)$ represents the left-continuous survival function estimated at time $t$ using the pooled survival time data from both groups using KM method.

FH test uses the weight  
\begin{equation*}
w_t(p, q) = \hat{S}(t-)^p \left(1 - \hat{S}(t-)\right)^q,
\end{equation*}
for $p \geq 0$ and $q \geq 0$. 
By appropriately selecting the hyperparameters $p$ and $q$, it is possible to assign greater weight to differences occurring in either the early or late stages of the study period.
In this study, we focus on survival scenarios such as late difference and early crossing survivals, where the survival curves of the two groups diverge significantly in the later stages of the trial. 
Therefore, the hyperparameters are set to $p=0$ and $q=1$.

Under the null and alternative hypotheses, 10,000 numerical simulations are repeated to calculate and evaluate the test size or power. 
Following the settings of \cite{pan2000two}, the numerical simulation of the two-sample WMST test is conducted under the PH and NPH assumptions. 
Similarly to the evaluation of the estimation methods, the generation of interval-censored data follows the method of \cite{zhang2020restricted}. 
However, in this subsection, all simulation settings are based on the default settings from the previous section, with only the distribution parameters for survival time $T$ being replaced. 
Figure \ref{sec3} illustrates the survival scenarios under the null and alternative hypotheses. 
In the PH scenario, the shape parameter $p$ is fixed and different scale parameters $\lambda$ are used for the Weibull distribution, resulting in a constant hazard ratio between the two groups of $\lambda_1^p/\lambda_0$, independent of time. 
For the piecewise-linear hazard function, both PH and NPH scenarios are considered. 
In the NPH case, the scenarios are classified as early difference, late difference, crossing hazards, and crossing survivals. 
The distributions or hazard functions of survival times under each scenario for both the null and alternative hypotheses are as follows.
\setlist[enumerate,1]{align=right,itemindent=0em}
\begin{itemize}[leftmargin=5mm]
	\item Null hypotheses
	\begin{itemize}[leftmargin=5mm]
		\item PH
		\begin{itemize}[leftmargin=5mm]
			\item Weibull
			\begin{enumerate}[label=(\roman*)]
				\item $T_0,T_1\sim \mathrm{Weibull}(\lambda=1,p=1)$
				\item $T_0,T_1\sim \mathrm{Weibull}(\lambda=1,p=0.5)$
				\item $T_0,T_1\sim \mathrm{Weibull}(\lambda=1,p=2)$
			\end{enumerate}
			\item Piecewise-linear hazard
			\begin{enumerate}[start=4,label=(\roman*)]
				\item $h_0(t)=h_1(t)=2$ ($t\in [0,0.5)$); $h_0(t)=h_1(t)=1$ ~($t\in [0.5,1]$)
				\item $h_0(t)=h_1(t)=1$ ($t\in [0,0.5)$); $h_0(t)=h_1(t)=2t$ ($t\in [0.5,1]$)
			\end{enumerate}
		\end{itemize}
	\end{itemize}
	\item Alternative hypotheses
	\begin{itemize}[leftmargin=5mm]
		\item PH
		\begin{itemize}[leftmargin=5mm]
			\item Weibull
			\begin{enumerate}[start=6,label=(\roman*)]
				\item $T_0\sim \mathrm{Weibull}(\lambda=0.5,p=1),~~~~T_1\sim \mathrm{Weibull}(\lambda=1,p=1)$
				\item $T_0\sim \mathrm{Weibull}(\lambda=0.5,p=0.5),~~T_1\sim \mathrm{Weibull}(\lambda=1,p=0.5)$
				\item $T_0\sim \mathrm{Weibull}(\lambda=0.75,p=2),~~~T_1\sim \mathrm{Weibull}(\lambda=1,p=2)$
			\end{enumerate}
			\item Piecewise-linear hazard
			\begin{enumerate}[start=9,label=(\roman*)]
				\item $h_0(t)=3,~~~h_1(t)=2$ ($t\in [0,0.5)$); $h_0(t)=1.5,~h_1(t)=1$ ~($t\in [0.5,1]$)
				\item $h_0(t)=1.5,~h_1(t)=1$ ($t\in [0,0.5)$); $h_0(t)=3t,~~h_1(t)=2t$ ($t\in [0.5,1]$)
			\end{enumerate}
		\end{itemize}
		\item NPH
		\begin{itemize}
			\item Early difference
			\begin{enumerate}[start=11,label=(\roman*)]
				\item $h_0(t)=1.75$, $h_1(t)=3t+0.25$ ($t\in [0,0.5)$); $h_0(t)=h_1(t)=t+1.25$ ($t\in [0.5,1]$)
			\end{enumerate}
			\item Late difference
			\begin{enumerate}[start=12,label=(\roman*)]
				\item $h_0(t)=h_1(t)=2$ ($t\in [0,0.2)$); $h_0(t)=4t+1.2$, $h_1(t)=-2t+2.4$ ($t\in [0.2,1]$)
			\end{enumerate}
			\item Crossing hazards
			\begin{enumerate}[start=13,label=(\roman*)]
				\item $h_0(t)=-1.5t+2$, $h_1(t)=1.5t+0.5$ ($t\in [0,1]$)
				\item $h_0(t)=1.5$, $h_1(t)=t+0.5$ ($t\in [0,0.5)$; $h_0(t)=0.5$, $h_1(t)=t+0.5$ ($t\in [0.5,1]$)
			\end{enumerate}
			\item Crossing survivals
			\begin{enumerate}[start=15,label=(\roman*)]
				\item ~Early: $h_0(t)=10t+1$, $h_1(t)=-10t+3$ ($t\in [0,0.2)$); $h_0(t)=3$, $h_1(t)=1$ ($t\in [0.2,1]$)
				\item Middle: $h_0(t)=1$, $h_1(t)=2$ ($t\in [0,0.25)$); $h_0(t)=3$, $h_1(t)=2$ ($t\in [0.25,1]$)
				\item ~~Late: $h_0(t)=2.5t+1$, $h_1(t)=-2.5t+3$, ($t\in [0,0.8)$); $h_0(t)=3$, $h_1(t)=1$ ($t\in [0.8,1]$)
			\end{enumerate}
		\end{itemize}
	\end{itemize}
\end{itemize}

\begin{figure}[H]
\centering
\includegraphics[width=\columnwidth]{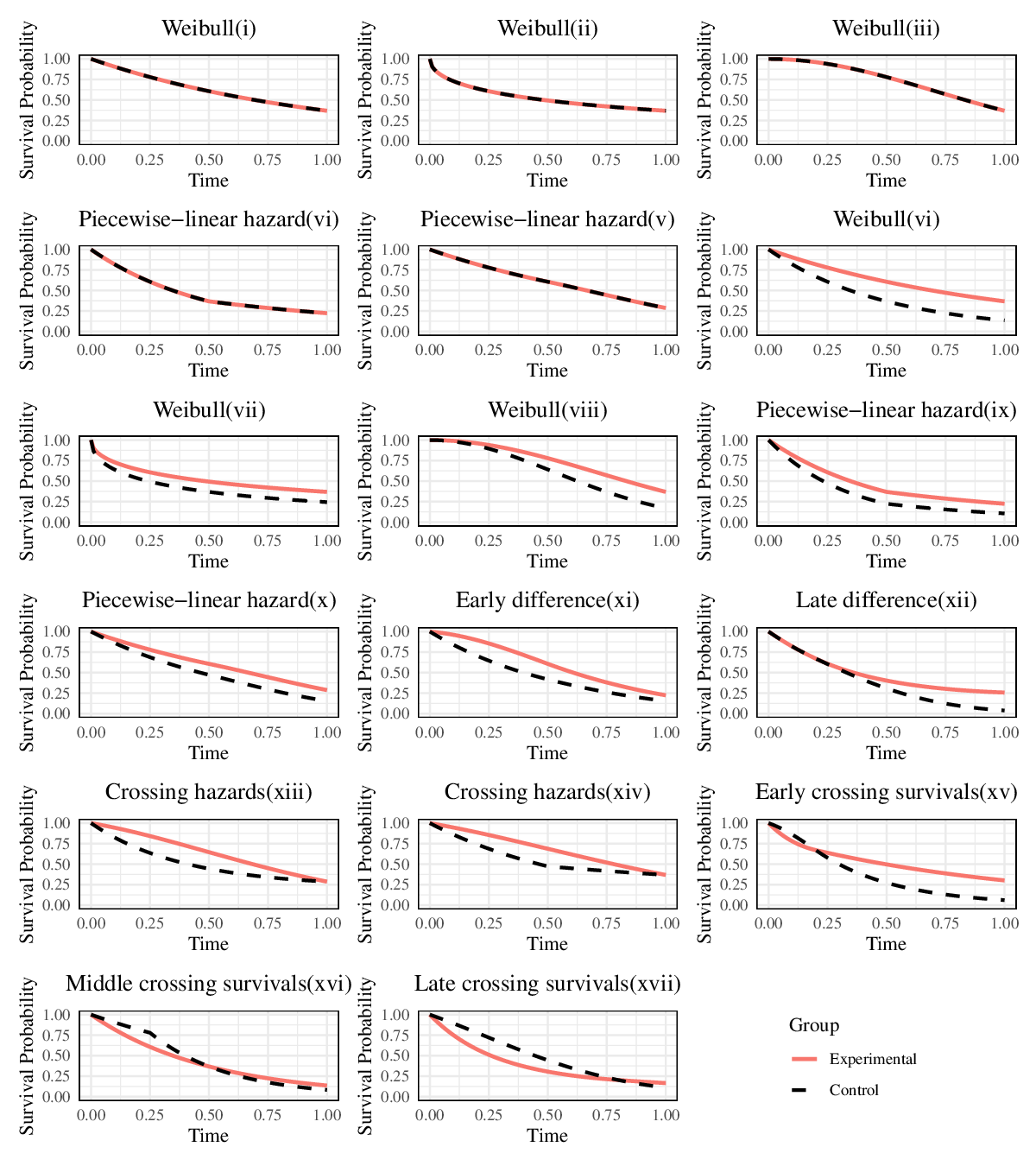}
\caption{The survival time scenarios used in the numerical simulations for hypothesis testing. $H_0$:(i)-(v), $H_1({\rm PH})$:(vi)-(x), $H_1({\rm NPH})$:(xi)-(xvii).}
\label{fig3}
\end{figure}

\begin{sidewaystable*}[!htbp]%
\begin{center}
\caption{Power for WMST tests considering $\tau_0$, the crossing point ($x$) of survival curves, and the difference ($\delta$) in RMST under the late difference scenario, compared to three existing methods: RMST, Log-rank, and FH.}
\label{tab2}
\begin{tabular*}{\textwidth}{@{\extracolsep\fill}lccccccccccccccc@{}}
\toprule
&\multicolumn{3}{c}{\textbf{RMST}}&\multicolumn{3}{c}{\textbf{WMST(}$\bm{\tau_0=0.25}$\textbf{)}}&\multicolumn{3}{c}{\textbf{WMST(}$\bm{\tau_0=0.50}$\textbf{)}}&\textbf{Log-rank}&\textbf{FH}$^{\rm a}$\\ \cmidrule(rl){2-4}\cmidrule(rl){5-7}\cmidrule(rl){8-10}\cmidrule(rl){11-11}\cmidrule(rl){12-12}
&\multicolumn{2}{c}{\raisebox{-0.1em}{\textbf{Diff}}}&&\multicolumn{2}{c}{\raisebox{-0.1em}{\textbf{Diff}}}&&\multicolumn{2}{c}{\raisebox{-0.1em}{\textbf{Diff}}}&&&\\ \cdashline{2-3}[3pt/2pt]\cdashline{5-6}[3pt/2pt]\cdashline{8-9}[3pt/2pt]
\raisebox{-0.3em}{\textbf{Distribution}}&\raisebox{-0.3em}{\textbf{True}}&\raisebox{-0.3em}{\textbf{Est.}}&&\raisebox{-0.3em}{\textbf{True}}&\raisebox{-0.3em}{\textbf{Est.}}&\raisebox{-0.3em}&\raisebox{-0.3em}{\textbf{True}}&\raisebox{-0.3em}{\textbf{Est.}}\\
\midrule
Under $H_0$&&&\textbf{size}&&&\textbf{size}&&&\textbf{size}&\textbf{size}&\textbf{size}\\ \cmidrule(r){1-1}
Weibull (i)&$0.0000$&$0.0000$&0.0540&0.0000&$0.0001$&0.0541&0.0000&0.0001&0.0540&0.0521&0.0521\\ \hdashline
Weibull (ii)&$0.0000$&$0.0002$&0.0514&0.0000&$0.0002$&0.0516&0.0000&0.0002&0.0535&0.0518&0.0512\\ \hdashline
Weibull (iii)&$0.0000$&$0.0000$&0.0530&0.0000&$0.0000$&0.0523&0.0000&0.0001&0.0517&0.0504&0.0495\\ \hdashline
Piecewise-linear hazard (iv)&$0.0000$&$0.0000$&0.0514&0.0000&$0.0000$&0.0502&0.0000&-0.0001&0.0509&0.0497&0.0509\\ \hdashline
Piecewise-linear hazard (v)&$0.0000$&$0.0003$&0.0542&0.0000&$0.0002$&0.0541&0.0000&0.0002&0.0535&0.0532&0.0546\\
\midrule
Under $H_1$ (PH)&&&\textbf{power}&&&\textbf{power}&&&\textbf{power}&\textbf{power}&\textbf{power}\\ \cmidrule(r){1-1}
Weibull (vi)&$0.1998$&$0.1902$&0.9753&0.1753&$0.1686$&0.9775&0.1224&0.1166&0.9756&0.9786&0.9664\\ \hdashline
Weibull (vii)&$0.1154$&$0.1103$&0.5086&0.0933&$0.0913$&0.5085&0.0631&0.0613&0.4983&0.5091&0.4791\\ \hdashline
Weibull (viii)&$0.1216$&$0.1174$&0.8470&0.1178&$0.1137$&0.8577&0.0952&0.0911&0.8749&0.8753&0.8477\\ \hdashline
Piecewise-linear hazard (ix)&$0.1234$&$0.1159$&0.7147&0.1025&$0.0976$&0.7064&0.0663&0.0622&0.6686&0.7130&0.6328\\ \hdashline
Piecewise-linear hazard (x)&$0.1133$&$0.1081$&0.6242&0.1006&$0.0969$&0.6415&0.0716&0.0684&0.6489&0.6510&0.6040\\
\midrule
Under $H_1$ (NPH)&&&\textbf{power}&&&\textbf{power}&&&\textbf{power}&\textbf{power}&\textbf{power}\\ \cmidrule(r){1-1}
Early difference (xi)&$0.1469$&$0.1387$&0.8720&0.1143&$0.1103$&0.7840&0.0612&0.0590&0.5583&0.7236&0.3626\\ \hdashline
Late difference (xii)&$0.0974$&$0.0860$&0.5058&0.0973&$0.0857$&0.6440&0.0861&0.0737&0.8823&0.7105&0.9627\\ \hdashline
crossing hazards (xiii)&$0.1397$&$0.1324$&0.7710&0.1076&$0.1044$&0.6700&0.0538&0.0524&0.4005&0.5356&0.1755\\ \hdashline
crossing hazards (xiv)&$0.1259$&$0.1198$&0.6721&0.1017&$0.0984$&0.5957&0.0519&0.0506&0.3482&0.4129&0.1521\\ \hdashline
Early crossing survivals (xv)&$0.1550$&$0.1441$&0.8743&0.1644&$0.1508$&0.9732&0.1248&0.1134&0.9929&0.9496&0.9941\\ \hdashline
Middle crossing survivals (xvi)&$-0.0211$&$-0.0195$&0.0791&0.0034&$0.0010$&0.0499&0.0210&0.0182&0.1209&0.0508&0.1743\\ \hdashline
Late crossing survivals (xvii)&$-0.0964$&$-0.0906$&$0.5306$&-0.0601&$-0.0592$&0.3425&$0.0140$&-0.0148&0.0922&0.3160&0.0615\\
\bottomrule
\end{tabular*}
\begin{tablenotes}
\item This simulation is conducted at a significance level of $\alpha=0.05$, with $n=100$ per arm, based on 10,000 replications.
\item $^{\rm a}$ FH is the Fleming-Harington test. 
\end{tablenotes}
\end{center}
\end{sidewaystable*}

Table \ref{tab2} shows that under the null hypothesis $H_0$, the test sizes for the four tests are close to the significance level $\alpha = 0.05$, effectively controlling the Type I error. 
Under the alternative hypothesis $H_1$, in the PH scenarios, the log-rank test demonstrates superior power compared to the other tests in most scenarios. 
Under the assumption of PH, WMST outperforms FH($p=0, q=1$) in terms of power in all scenarios. 
In NPH scenarios, WMST shows higher power than RMST in late difference and early crossing survival scenarios. 

For survival scenarios where WMST demonstrates superior power compared to RMST (late difference and early crossing survivals), FH test exhibits even higher power than WMST. 
However, it is important to note that, under the PH, WMST consistently outperforms FH in terms of power. 
The test methods must be selected prior to the start of a trial. 
If clinicians assume the NPH before the trial and select FH as the testing method, but the actual survival time data satisfy the PH assumption, there is a concern that this could lead to a loss of power.

Moreover, not only under the PH but also in the NPH cases such as early difference and two types of crossing hazards, FH test exhibits significantly lower power compared to WMST, which is an undesirable characteristic. 
Additionally, under the PH, the log-rank test allows estimation of the hazard ratio by applying the proportional hazards model, enabling a quantitative assessment of the effect size between two groups associated with the test. 
In contrast, FH test does not provide a quantitative measure of the effect size between two groups, such as a difference or a ratio corresponding to the test, making its interpretation highly challenging.

In contrast, WMST provides the mean survival time within a specified window, which is easier to interpret compared to the hazard ratio. 
This advantage extends beyond clinical trials and enables even clinicians or patients without advanced statistical knowledge to make informed and confident decisions about treatment plans or drug choices based on a clear understanding.

\subsection{Altering choice of $\tau_0$, crossing point and difference of RMST} \label{choice of tau} 
The setting of $\tau_0$ is determined prior to the collection of trial data in actual clinical trials. 
In this study, numerical simulations are conducted to investigate how the pre-specified value of $\tau_0$ affects the power to detect differences when the divergence or crossing point of the two survival curves derived from the trial data occurs at different points. 
In addition, the impact of the true difference in RMST on power is examined. 
Focusing on the late difference and early crossing survivals scenarios, where WMST shows superior results to RMST in the previous subsection, the divergence or crossing point $x$ of the two survival curves is set at $\{0.10, 0.15, 0.20, 0.25, 0.30\}$, $\tau_0$ at $\{0.05, 0.10, 0.15, 0.20, 0.25, 0.30, 0.35, 0.40, 0.45, 0.50\}$, and the difference of RMST $\delta$ is set at $\{0.10, 0.15, 0.20\}$. 
For all $5 \times 10 \times 3$ combinations, 5,000 numerical simulations are conducted to calculate the power. 
As in the previous subsection, all other parameters are kept in their default settings. 

Table \ref{tab3} presents the results for the survival scenario with a late difference. 
Regardless of the value of $\tau_0$, WMST consistently demonstrates higher power than RMST. 
For WMST, even when there is a misalignment between the divergence point $x$ of the survival curves and $\tau_0$, power improved when the misalignment is in the backward direction. 
Even with a forward misalignment, WMST still shows higher power than RMST. 
Furthermore, as the true difference in RMST $\delta$ increased, both RMST and WMST shows improvements in power. 
When $x$ and $\tau_0$ are aligned, which means that the interval of WMST window started exactly from the divergence point, the power of WMST is slightly behind that of the log-rank test. 

\begin{sidewaystable*}[!htbp]%
\begin{center}
\caption{Power for WMST tests based on $\tau_0$, the crossing point ($x$) of survival curves and difference of RMST under the late difference scenario, compared to three existing methods: RMST, log-rank, and FH.}
\label{tab3}
\begin{tabular*}{\textwidth}{@{\extracolsep\fill}lccccccccccccccc@{}}
\toprule
&\textbf{RMST}&\multicolumn{10}{c}{\textbf{WMST}}&\textbf{Log-rank}&\textbf{FH}$^{\rm a}$\\ \cmidrule(rl){2-2}\cmidrule(rl){3-12}\cmidrule(rl){13-13}\cmidrule(rl){14-14}
&&\raisebox{-0.3em}{$\bm{\tau_0=0.05}$}&\raisebox{-0.3em}{$\bm{0.10}$}&\raisebox{-0.3em}{$\bm{0.15}$}&\raisebox{-0.3em}{$\bm{0.20}$}&\raisebox{-0.3em}{$\bm{0.25}$}&\raisebox{-0.3em}{$\bm{0.30}$}&\raisebox{-0.3em}{$\bm{0.35}$}&\raisebox{-0.3em}{$\bm{0.40}$}&\raisebox{-0.3em}{$\bm{0.45}$}&\raisebox{-0.3em}{$\bm{0.50}$}\\
\midrule
$x=0.10$&&&&&&&&&&&\\ \cmidrule{1-1}
$\delta=0.10$&$0.5398$&$0.5426$&0.5566&0.5922&$0.6274$&0.6764&0.7228&0.7782&0.8200&0.8568&0.8876&0.7404&0.9562\\ \hdashline
$\delta=0.15$&$0.8160$&$0.8190$&0.8358&0.8590&$0.8894$&0.9230&0.9494&0.9688&0.9824&0.9904&0.9930&0.9474&0.9990\\ \hdashline
$\delta=0.20$&$0.9462$&$0.9476$&0.9564&0.9678&$0.9788$&0.9882&0.9938&0.9970&0.9986&0.9992&0.9996&0.9930&1.0000\\
\midrule
$x=0.15$&&&&&&&&&&&\\ \cmidrule{1-1}
$\delta=0.10$&$0.5458$&$0.5474$&0.5654&0.5974&$0.6364$&0.6832&0.7338&0.7818&0.8224&0.8570&0.8924&0.7422&0.9558\\ \hdashline
$\delta=0.15$&$0.8104$&$0.8142$&0.8298&0.8550&$0.8860$&0.9192&0.9498&0.9670&0.9810&0.9884&0.9924&0.9940&0.9992\\ \hdashline
$\delta=0.20$&$0.9552$&$0.9566$&0.9636&0.9734&$0.9822$&0.9912&0.9950&0.9980&0.9988&0.9996&0.9996&0.9954&1.0000\\
\midrule
$x=0.20$&&&&&&&&&&&\\ \cmidrule{1-1}
$\delta=0.10$&$0.5464$&$0.5504$&0.5710&0.5992&$0.6408$&0.6858&0.7362&0.7872&0.8270&0.8616&0.8970&0.7448&0.9588\\ \hdashline
$\delta=0.15$&$0.8270$&$0.8306$&0.8458&0.8728&$0.8998$&0.9302&0.9556&0.9734&0.9858&0.9908&0.9940&0.9522&0.9992\\ \hdashline
$\delta=0.20$&$0.9604$&$0.9628$&0.9690&0.9768&$0.9848$&0.9924&0.9956&0.9984&0.9994&0.9996&1.0000&0.9958&1.0000\\
\midrule
$x=0.25$&&&&&&&&&&&\\ \cmidrule{1-1}
$\delta=0.10$&$0.5436$&$0.5462$&0.5674&0.5952&$0.6370$&0.6840&0.7362&0.7852&0.8258&0.8680&0.9024&0.7456&0.9638\\ \hdashline
$\delta=0.15$&$0.8308$&$0.8338$&0.8470&0.8756&$0.9014$&0.9340&0.9578&0.9752&0.9864&0.9912&0.9954&0.9544&0.9992\\ \hdashline
$\delta=0.20$&$0.9646$&$0.9654$&0.9692&0.9784&$0.9862$&0.9924&0.9962&0.9986&0.9994&0.9996&1.0000&0.9958&1.0000\\
\midrule
$x=0.30$&&&&&&&&&&&\\ \cmidrule{1-1}
$\delta=0.10$&$0.5438$&$0.5466$&0.5646&0.5940&$0.6342$&0.6852&0.7360&0.7892&0.8344&0.8796&0.9108&0.7550&0.9692\\ \hdashline
$\delta=0.15$&$0.8302$&$0.8324$&0.8488&0.8734&$0.9004$&0.9314&0.9592&0.9768&0.9884&0.9936&0.9962&0.9564&0.9994\\ \hdashline
$\delta=0.20$&$0.9648$&$0.9662$&0.9712&0.9800&$0.9894$&0.9934&0.9968&0.9986&0.9994&1.0000&1.0000&0.9972&1.0000\\
\bottomrule
\end{tabular*}
\begin{tablenotes}
\item This simulation is conducted at a significance level of $\alpha=0.05$, with $n=100$ per arm, based on 5,000 replications.
\item $^{\rm a}$ FH is the Fleming-Harington test. 
\end{tablenotes}
\end{center}
\end{sidewaystable*}

Table \ref{tab4} presents the results for the survival scenario of early crossing survivals. 
Similarly to Table 3, WMST consistently demonstrates higher power than RMST regardless of the value of $\tau_0$. 
For WMST, even when there is a misalignment between the divergence point $x$ of the survival curves and $\tau_0$, power improved when the misalignment is in the backward direction. 
Even with a forward misalignment, WMST always shows higher power than RMST. 
Furthermore, as the true difference in RMST $\delta$ increases, both RMST and WMST exhibits improvements in power. 
When $x$ and $\tau_0$ are aligned and $x = \tau_0 \geq 0.15$, WMST outperforms the log-rank test.

\begin{sidewaystable*}[!htbp]%
\begin{center}
\caption{Power for WMST tests based on $\tau_0$, the crossing point ($x$) of survival curves and difference of RMST under the early crossing survivals scenario, compared to three existing methods: RMST, log-rank, and FH.}
\label{tab4}
\begin{tabular*}{\textwidth}{@{\extracolsep\fill}lccccccccccccccc@{}}
\toprule
&\textbf{RMST}&\multicolumn{10}{c}{\textbf{WMST}}&\textbf{Log-rank}&\textbf{FH}$^{\rm a}$\\ \cmidrule(rl){2-2}\cmidrule(rl){3-12}\cmidrule(rl){13-13}\cmidrule(rl){14-14}
&&\raisebox{-0.3em}{$\bm{\tau_0=0.05}$}&\raisebox{-0.3em}{$\bm{0.10}$}&\raisebox{-0.3em}{$\bm{0.15}$}&\raisebox{-0.3em}{$\bm{0.20}$}&\raisebox{-0.3em}{$\bm{0.25}$}&\raisebox{-0.3em}{$\bm{0.30}$}&\raisebox{-0.3em}{$\bm{0.35}$}&\raisebox{-0.3em}{$\bm{0.40}$}&\raisebox{-0.3em}{$\bm{0.45}$}&\raisebox{-0.3em}{$\bm{0.50}$}\\
\midrule
$x=0.10$&&&&&&&&&&&\\ \cmidrule(r){1-1}
$\delta=0.10$&$0.6012$&$0.6074$&0.6320&0.6554&$0.6794$&0.6912&0.7046&0.7114&0.7134&0.7112&0.7086&0.6422&0.6692\\ \hdashline
$\delta=0.15$&$0.8634$&$0.8688$&0.8866&0.9064&$0.9242$&0.9356&0.9428&0.9486&0.9500&0.9514&0.9504&0.9032&0.9350\\ \hdashline
$\delta=0.20$&$0.9756$&$0.9762$&0.9818&0.9858&$0.9914$&0.9930&0.9952&0.9960&0.9960&0.9962&0.9970&0.9858&0.9958\\
\midrule
$x=0.15$&&&&&&&&&&&\\ \cmidrule(r){1-1}
$\delta=0.10$&$0.5834$&$0.5950$&0.6372&0.6772&$0.7124$&0.7378&0.7600&0.7742&0.7856&0.7888&0.7930&0.6586&0.7624\\ \hdashline
$\delta=0.15$&$0.8678$&$0.8732$&0.8956&0.9202&$0.9400$&0.9550&0.9646&0.9690&0.9716&0.9744&0.9756&0.9190&0.9698\\ \hdashline
$\delta=0.20$&$0.9764$&$0.9784$&0.9846&0.9898&$0.9934$&0.9964&0.9972&0.9974&0.9976&0.9984&0.9984&0.9902&0.9992\\
\midrule
$x=0.20$&&&&&&&&&&&\\ \cmidrule(r){1-1}
$\delta=0.10$&$0.5670$&$0.5776$&0.6264&0.6848&$0.7362$&0.7756&0.8094&0.8348&0.8482&0.8546&0.8624&0.6946&0.8528\\ \hdashline
$\delta=0.15$&$0.8614$&$0.8674$&0.8936&0.9248&$0.9504$&0.9676&0.9784&0.9838&0.9860&0.9892&0.9906&0.9358&0.9900\\ \hdashline
$\delta=0.20$&$0.9766$&$0.9772$&0.9834&0.9918&$0.9958$&0.9978&0.9984&0.9984&0.9990&0.9994&0.9998&0.9942&0.9998\\
\midrule
$x=0.25$&&&&&&&&&&&\\ \cmidrule(r){1-1}
$\delta=0.10$&$0.7726$&$0.7818$&0.8272&0.8750&$0.9220$&0.9508&0.9682&0.9758&0.9802&0.9816&0.9828&0.9084&0.9780\\ \hdashline
$\delta=0.15$&$0.8504$&$0.8582$&0.8926&0.9298&$0.9586$&0.9770&0.9864&0.9914&0.9946&0.9962&0.9972&0.9572&0.9974\\ \hdashline
$\delta=0.20$&$0.9728$&$0.9746$&0.9824&0.9918&$0.9966$&0.9984&0.9992&0.9996&0.9998&1.0000&1.0000&0.9966&1.0000\\
\midrule
$x=0.30$&&&&&&&&&&&\\ \cmidrule(r){1-1}
$\delta=0.10$&$0.5542$&$0.5694$&0.6360&0.7238&$0.8062$&0.8724&0.9218&0.9492&0.9654&0.9758&0.9800&0.8318&0.9844\\ \hdashline
$\delta=0.15$&$0.8472$&$0.8552$&0.8948&0.9376&$0.9680$&0.9852&0.9928&0.9974&0.9980&0.9988&0.9990&0.9788&0.9998\\ \hdashline
$\delta=0.20$&$0.9688$&$0.9710$&0.9806&0.9920&$0.9966$&0.9990&0.9998&1.0000&1.0000&1.0000&1.0000&0.9982&1.0000\\
\bottomrule
\end{tabular*}
\begin{tablenotes}
\item This simulation was conducted at a significance level of $\alpha=0.05$, with $n=100$ per arm, based on 5,000 replications.
\item $^{\rm a}$ FH is the Fleming-Harington test. 
\end{tablenotes}
\end{center}
\end{sidewaystable*}

\section{Application} \label{sec4}
\begin{figure*}[!htbp]
\centering
\includegraphics[width=\columnwidth]{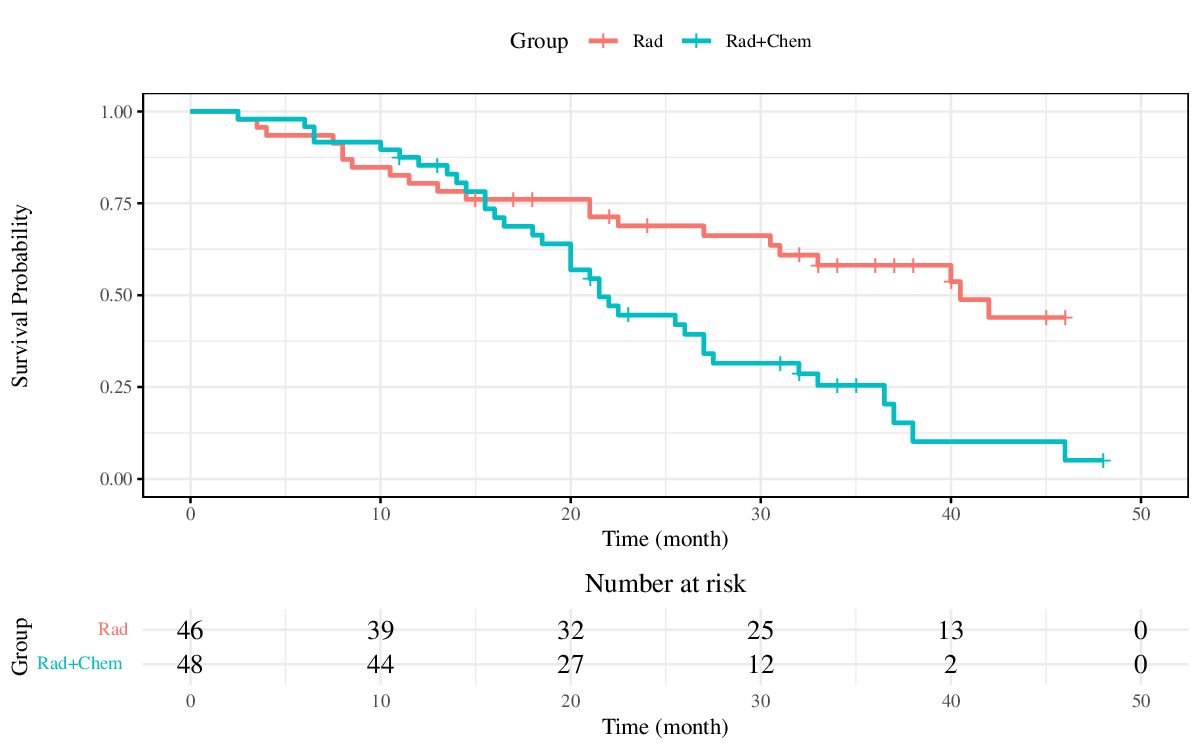}
\caption{KM curves estimated by using the mid-point imputation method on interval-censored data in the BCOS dataset, where `Rad' represents the radiotherapy group and `Rad + Chem' represents the group that received radiotherapy with adjuvant chemotherapy.}
\label{fig4}
\end{figure*}

This section demonstrates the results of applying the WMST method to real data with interval censoring. 
Similarly to the previous section, the interval-censored data are imputed using the mid-point imputation method, after which the survival function is estimated with KM method. 
WMST is then estimated and tested, and the results are compared with existing testing methods. 
The real data used in this study are from the `bcos` dataset available in the \texttt{interval} package of \texttt{R} software \citep{finkelstein1985semiparametric}. 
The \texttt{bcos} dataset contains interval-censored survival times for 46 patients who underwent radiation therapy and 48 patients who received a combination of radiation therapy and chemotherapy, with the survival endpoint defined as cosmetic deterioration. 
The survival curves estimated with KM method after the mid-point imputation are shown in Figure \ref{fig4}. 
The two survival curves intersect around 15 months, corresponding to the early crossing survivals scenario discussed in the previous chapter. 

The $p$-values for each test compared in the previous section are as follows: $0.0091$ (RMST), $0.0010$ (WMST($\tau_0 = 15$)), $0.0011$ (log-rank), and $0.0001$ (FH). 
As in the previous section, WMST is tested by shifting $\tau_0$ around the intersection point of the survival curves ($\tau_0 \in \{12.5, 17.5\}$). 
The resulting $p$-values are $0.0021$ (WMST($\tau_0 = 12.5$)) and $0.0004$ (WMST($\tau_0 = 17.5$)).
As a result, the $p$-values for all tests are below $0.05$, indicating a significant difference in survival rates between the radiation-only group and the combination therapy group. 
It is well known that under the NPH, the log-rank test and FH test cannot provide the clinically meaningful interpretation of the difference between the survival curves of the two groups. 
In contrast, for RMST, the difference between the groups is estimated to be $7.06$ (95\% CI [$1.76, 12.37$]), and for WMST, the difference between the groups is estimated to be $7.53$ (95\% CI [$3.06, 12.00$]). 
Furthermore, in this case, as the survival curves crossed early, the $p$-value for WMST is lower than that for RMST.

\section{Conclusion} \label{sec5}
In recent years, the NPH survival scenarios have been steadily increasing in cancer clinical trials. 
In particular, cancer immunotherapies, which have gained significant attention, often exhibit late differences and early crossing survivals scenarios.
Furthermore, in phase III trials targeting advanced cancers, the PFS is increasingly being adopted as a co-primary endpoint alongside the OS. 
In response to these trends, we propose a class of methods for the estimation and testing of WMST designed for interval-censored data and conduct extensive numerical simulations to compare their performance with conventional methods. 

Through numerical simulations, WMST estimation method that uses mid-point imputation followed by KM estimation (mid-point $+$ KM) demonstrates estimation accuracy that is almost equal to that of WMST estimation method with Turnbull, which is the NPMLE. 
Unlike the frequently used right-point $+$ KM method in clinical practice, WMST estimation methods based on the mid-point $+$ KM and Turnbull are shown to be robust to both the proportion of interval-censored cases and the interval widths of the interval-censored data.
Considering these findings, we establish WMST estimation based on mid-point $+$ KM as the standard approach, given its ability to explicitly express standard errors and its ease of implementation. 
Numerical simulations fot the hypothesis testing show that even in the presence of interval-censored data, WMST maintains higher power than FH($p=0, q=1$) under the PH survival scenarios. 
Furthermore, WMST outperforms RMST in terms of power under late difference and early crossing survivals sceanario, which are frequently observed in cancer immunotherapy trials. 
Furthermore, in the late difference and early crossing survivals scenarios, the impact of selection $\tau_0$ is investigated. 
It is found that when the value of $\tau_0$ is shifted backward relative to the divergence point or the crossing point of the two curves, power increased. 
Even when $\tau_0$ is moved forward, WMST maintains higher power than RMST for interval-censored data. 
Our findings highlight that the proposed method is anticipated to be a more practical approach for assessing PFS in cancer immunotherapy trials, particularly under late difference and early crossing survival scenarios. 

In this study, we employ the mid-point imputation for handling interval-censored data. 
However, the mid-point imputation assigns a single value to all subjects whose disease progression (DP) is observed within the same interval. 
This means that all DP events within the same interval occur on the same day, which is unrealistic. 
To address this issue, \cite{nakagawa2024improvement} propose an enhanced mid-point imputation, and their numerical simulations demonstrate that the enhanced method provides more accurate estimates of the survival function compared to the standard mid-point imputation. 
While applying the enhanced method in our study could lead to better WMST estimation accuracy and statistical power, a rigorous evaluation of these potential improvements remains an open question for future research.

\bibliographystyle{apalike} 
\bibliography{references.bib}

\end{document}